\documentclass[11pt]{article}

\usepackage[a4paper,margin=1in]{geometry}
\usepackage{graphicx}
\usepackage{times}
\usepackage{amssymb,amsmath}
\usepackage[authoryear,round]{natbib}
\usepackage[pagebackref=true,breaklinks=true,hidelinks,hyperfootnotes=false]{hyperref}
\usepackage{xurl}
\usepackage{setspace}
\usepackage{authblk}

\title{Debris Evolution from Spacecraft Fragmentation in Earth-Moon Distant Retrograde Orbits}

\author[1,2]{Yuyan Wu}
\author[1]{Peng Shu\thanks{Corresponding author: \href{mailto:shupeng@ynao.ac.cn}{shupeng@ynao.ac.cn}}}
\author[1]{Yuqiang Li}
\affil[1]{Yunnan Observatories, Chinese Academy of Sciences, Kunming 650011, China}
\affil[2]{University of Chinese Academy of Sciences, Beijing 100049, China}
\date{}

\begin{document}

\maketitle

\begin{abstract}
With the rapid surge in lunar exploration and the planned expansion of cislunar infrastructure, cislunar space has become a strategic focal point for global aerospace activities. This proliferation of spacecraft heightens the risk of fragmentation events, such as unintended explosions or orbital collisions which serve as the primary source of hazardous orbital debris. Given the potential threat these fragments pose to mission safety and long-term orbital sustainability, it is imperative to investigate their dynamical behavior within the Earth-Moon system. This study evaluates the dispersion of debris clouds following potential breakup events on Distant Retrograde Orbits (DROs) over a 30-day propagation period. The Circular Restricted Three-Body Problem (CR3BP) model is used to construct the reference orbits, while the NASA Standard Breakup Model is applied to simulate fragment generation at multiple locations along three DROs of varying sizes. These fragments are then propagated using the Bicircular Restricted Four-Body Problem (BCR4BP) for 30 days. To account for the variability of these events, multiple initial positions along each orbit are analyzed to capture a comprehensive range of post-explosion scenarios. Our analysis quantifies the fate of fragments within this window, specifically focusing on the escape mechanisms and the percentages of debris that either depart from the Earth-Moon gravitational sphere of influence or impact the lunar surface. Furthermore, we introduce an analytical approach to assess the potential collision risk to resident space objects operating within the vicinity of the parent orbit. The results provide insights into debris evolution and offer a foundation for developing safety guidelines for future cislunar activities.
\end{abstract}


\section{Introduction}

With the rapid advancement of the global aerospace industry \citep{articleMelamed}, the scope of human activity has expanded from traditional Low Earth Orbit (LEO) and Geostationary Earth Orbit (GEO) to the broader cislunar space. Cislunar space, defined as the region extending beyond the geosynchronous belt to the Moon's sphere of influence, is regarded as a strategic "high ground" for space operations. It offers significant value in terms of positioning, logistics, and resource utilization. In recent years, international missions in this region have surged, including the China National Space Administration's (CNSA) Chang'e lunar exploration series \citep{articleli}. The Indian Space Research Organisation's (ISRO) Chandrayaan-3\footnote{\url{https://www.isro.gov.in/Chandrayaan3_Details.html}}, and NASA's pathfinder mission CAPSTONE have further advanced cislunar mission capabilities. CAPSTONE successfully operated in the Earth–Moon L2 Near Rectilinear Halo Orbit (NRHO)\footnote{\url{https://www.nasa.gov/smallsat-institute/community-of-practice/an-overview-and-status-of-the-capstone-mission/}}. This orbit has been selected for the Gateway lunar outpost and serves as a cornerstone of the Artemis program’s lunar exploration architecture\footnote{\url{https://www.nasa.gov/image-article/artemis-ii-map-2/}}.

However, the proliferation of space activities inevitably escalates the risk of catastrophic fragmentation events. Such incidents may arise from collisions between spacecraft, passivation failures leading to propellant tank ruptures, battery explosions, or the natural degradation of aging structural components as discussed in technical guidelines \citep{SchonbergHull2023Passivation}. In LEO, fragmentation debris already constitutes the primary source of space objects, accounting for over 50\% of the cataloged population. In the context of cislunar space, even a single isolated fragmentation event has the potential to generate a vast number of debris clouds, posing significant challenges to an orbital domain already governed by chaotic dynamics. The destructive potential of such events is well documented. For example, the oxygen tank explosion during the Apollo 13 mission in 1970 generated a debris field that not only threatened the crew's safety but also impaired navigation by obscuring stellar references. Furthermore, the collision between Cosmos 2251 and Iridium 33 in LEO demonstrated that debris from a single event can disperse over vast volumes, potentially triggering cascading collisions.

Unlike LEO, where atmospheric drag provides a natural cleansing mechanism \citep{Andrenucci_etal_2011_ESA_DebrisRemoval}, the lunar environment lacks an atmosphere, meaning debris cannot decay and re-enter naturally. Studies indicate that objects in Near-Rectilinear Halo Orbits (NRHOs) can persist for decades \citep{2023AdSpR..72.1550G}. This long-term persistence in cislunar space increases the likelihood of debris accumulation \citep{2023amos.conf...20B, inproceedingsguard}. Such accumulation is particularly significant in dynamically sensitive regions, including the Lagrange points (L1 and L2) and unstable invariant manifolds. In these regions, fragments can remain in circulation for extended periods \citep{articleBlack}. This prolonged residence increases the collision risk to future lunar stations, surface infrastructure, and transfer trajectories.

Among various cislunar orbits, DROs have garnered significant attention due to their unique stability \citep{inproceedingsBezrouk}, low insertion energy requirements \citep{2025AsDyn...9..165W}, and strategic utility as staging points for deep space exploration\footnote{Chinese Academy of Sciences, 2025, China Achieves Its 1st Lunar-distance Satellite Laser Ranging, \url{https://english.cas.cn/newsroom/cas_media/202504/t20250427_1042108.shtml}, accessed 2025-10-28}. DROs are considered ideal locations for long-term infrastructure such as space stations and fuel depots. Nevertheless, assessing fragmentation risks in these orbits is fraught with difficulty. The challenge lies in the immense parameter space; even within the simplified CR3BP model, which ignores solar radiation pressure and gravitational perturbations from the Sun and Jupiter, the variability is substantial. Fragmentation events differ in type, intensity, parent mass, and the number and energy of generated fragments \citep{2001AdSpR..28.1377J}. Additionally, the dynamical evolution depends on the specific trajectory or orbit family and the precise location of the event along the trajectory \citep{2023JSpRo..60..668B}. Consequently, research has often been limited to isolated case studies, such as events on specific Lyapunov or Halo orbits \citep{Nishiguchi_etal_2025}.

Existing literature on cislunar debris has largely focused on Halo orbits or NRHOs near libration points, identifying mechanisms like "self-cleaning" \citep{2023amos.conf...20B, 2023amos.conf...79B}. However, for the highly stable DRO family, the evolutionary behavior of debris—specifically how fragments are retained or decay over time and the dependence of this behavior on the initial orbital phase—lacks systematic quantification and mechanistic understanding. This study aims to bridge this gap by employing theoretical modeling and large-scale numerical simulations to elucidate the dynamical evolution of debris resulting from spacecraft breakups on DROs in the Earth-Moon system.

This research contributes to the field by providing a quantitative risk assessment for DRO debris environments, analyzing how natural clearance efficiency and 30-day retention risks vary with the characteristics of the parent DRO. A key focus is on revealing phase-dependent evolutionary patterns, demonstrating how the orbital phase at the moment of breakup governs transport pathways. Notably, events occurring on the lunar far side tend to produce debris trapped within the system, whereas near-Earth breakups may exhibit distinct escape characteristics. This study analyzes the fate of DRO fragments, specifically whether they impact the Moon, escape the system, or stay in the orbit vicinity — to provide evidence-based recommendations for debris mitigation. These findings support decision-making for future infrastructure, such as lunar relay satellites and the Lunar Gateway. The study utilizes the NASA Standard Breakup Model (SBM) to generate high-fidelity initial debris states and propagates trajectories using both CR3BP and BCR4BP models.

\section{Orbital Dynamics Model}
\label{sec:DATA-AND-METHODS}

This study employs two primary dynamical models: the Circular Restricted Three-Body Problem (CR3BP) and the Bi-circular Restricted Four-Body Problem (BCR4BP). 
Following the approach of \citet{Henon1969Hill}, the CR3BP is used to construct reference DRO families and determine the breakup locations sampled along each orbit. The BCR4BP extends this framework by incorporating solar perturbations to achieve higher-fidelity trajectory propagation. 
As demonstrated in \citet{2023JSpRo..60..668B}, the method of propagating debris fragments in a higher-fidelity model has been widely adopted.

\subsection{Circular Restricted Three Body Problem}
The CR3BP is a fundamental model for studying orbital dynamics in a system with two dominant masses, such as Earth and the Moon. In this model, the primaries (Earth and Moon) orbit their common barycenter in circular Keplerian trajectories, and the third body (the spacecraft) has negligible mass and does not influence the primaries. These assumptions yield a simplified, computationally efficient dynamical framework.

We use a rotating (synodic) reference frame centered at the Earth–Moon barycenter. In this frame, the x-axis connects the two primaries, the z-axis aligns with their orbital angular momentum, and the y-axis completes a right-handed system. In this frame, the primaries are fixed on the x-axis, yielding a time-invariant formulation. The third body’s motion follows from a pseudo-potential U* that incorporates both primaries’ gravitational potentials and the centrifugal potential of the rotating frame:
\begin{align}
\ddot{x} - 2\dot{y} = \frac{\partial U^{*}_{\text{CR3BP}}}{\partial x},
\ddot{y} + 2\dot{x} = \frac{\partial U^{*}_{\text{CR3BP}}}{\partial y},
\ddot{z} = \frac{\partial U^{*}_{\text{CR3BP}}}{\partial z} . \label{eq:cr3bp}
\end{align}
Equation~(\ref{eq:cr3bp}) follows the standard formulation of the CR3BP equations of motion in the rotating frame \citep{Szebehely1967book,Howell2018}.

Here, $\mu = M_2/(M_1+M_2)$ (about 0.012 for Earth–Moon system) is the mass parameter of the system, a dimensionless quantity between 0 and 0.5. The dynamics are governed by the pseudo-potential $U^*$, which incorporates both the gravitational potential of the two primaries and the centrifugal potential arising from the rotation of the reference frame, defined as: $$U^* = \frac{1}{2}(x^2 + y^2) + \frac{1-\mu}{r_1} + \frac{\mu}{r_2}$$The distances from the third body to the larger primary ($M_1$, located at $x=-\mu$) and the smaller primary ($M_2$, located at $x=1-\mu$) are $r_1 = \sqrt{(x+\mu)^2 + y^2 + z^2}$ and $r_2 = \sqrt{(x-1+\mu)^2 + y^2 + z^2}$, respectively. The terms $2\dot{y}$ and $2\dot{x}$ represent the Coriolis acceleration, a characteristic effect of the rotating frame.  

The CR3BP exhibits five equilibrium solutions known as Lagrange points, where gravitational and centrifugal forces balance. These points structure the system’s dynamics. L1, L2, and L3 lie along the Earth–Moon line (x-axis) and are linearly unstable, serving as dynamic gateways between regions. L4 and L5 form equilateral triangles with the primaries and are stable (for the Earth–Moon mass ratio) under small perturbations.

 \subsection{Bicircular Restricted Four-Body Problem}

While the CR3BP captures Earth–Moon dynamics, it ignores other perturbing masses, notably the Sun. For long-duration cislunar missions or orbits sensitive to solar gravity, this approximation fails. BCR4BP addresses this by adding the Sun as a third primary on a fixed circular orbit (around the Earth–Moon barycenter in the synodic frame). This yields an intermediate-fidelity model between the CR3BP and a full ephemeris propagator \citep{Simo1995,Negri2020}.

In the BCR4BP, Earth and Moon (the primary pair) still orbit their common barycenter in circular paths (as in CR3BP). The Sun is introduced as a third massive body moving on a circular orbit about that same barycenter. While this assumes the Earth–Moon barycenter is fixed while the Sun moves, which is an approximation of physical reality, it is a common approximation. The spacecraft remains massless, influenced by the gravitational forces of Earth, Moon, and Sun, plus the Coriolis and centrifugal forces of the rotating frame.

The equations of motion follow directly from augmenting the CR3BP dynamics with the Sun’s direct gravitational attraction and an indirect term accounting for the acceleration of the rotating frame’s origin induced by the Sun:
\begin{align}
\ddot{x} - 2\dot{y} &= \frac{\partial U^{*}_{\text{CR3BP}}}{\partial x} + \mu_s \left( \frac{ x_s - x }{ {r_{s}}^3 } - \frac{ x_s }{ {R_{s}}^3 } \right), \label{eq:bcr4bp_x} \\
\ddot{y} + 2\dot{x} &= \frac{\partial U^{*}_{\text{CR3BP}}}{\partial y} + \mu_s \left( \frac{ y_s - y }{ {r_{s}}^3 } - \frac{ y_s }{ {R_{s}}^3 } \right), \label{eq:bcr4bp_y} \\
\ddot{z} &= \frac{\partial U^{*}_{\text{CR3BP}}}{\partial z} + \mu_s \left( \frac{ z_s - z }{ {r_{s}}^3 } \right). \label{eq:bcr4bp_z}
\end{align}
 Here, $\mu_s$ is the nondimensional mass parameter of the Sun, $R_s$ is its constant distance from the Earth–Moon barycenter, and
  $
  r_s = \sqrt{(x - x_s)^2 + (y - y_s)^2 + (z - z_s)^2}
  $
  denotes the instantaneous distance between the spacecraft and the Sun. The Sun’s position in the synodic frame evolves as
  
  $x_s = R_s\cos\theta_s$,
  $y_s = R_s\sin\theta_s$,
  $z_s = 0$,
  with a phase angle
  $\theta_s = \theta_{s0} + n_s t$,
  where $n_s$ is the angular velocity of the Sun relative to the Earth–Moon rotating frame.

The inclusion of the Sun introduces explicit time dependence into the equations of motion, rendering the BCR4BP a non-autonomous dynamical system. Despite its simplified assumptions, the BCR4BP strikes an effective balance between model fidelity and computational efficiency. It captures the leading-order solar perturbation absent in the CR3BP while avoiding the complexity of full ephemeris models. 
\begin{table} 
\centering
\caption{Fragmentation data.}
\label{tab:fragmentation_data}
\begin{tabular}{ll}
\hline
\textbf{Event type}         & Explosion  \\
\textbf{Body type}          & Satellite  \\
\textbf{Mass}               & 1000 kg    \\
\textbf{Min. size propagated} & 1 cm       \\
 
\hline
\textbf{Tot. objects}       & 9409      \\
\hline
\end{tabular}
\end{table}

\subsection{Finite-Time Lyapunov Exponent}
\label{subsection:FTLE}

The Finite-Time Lyapunov Exponent (FTLE) is a standard diagnostic for
quantifying the sensitivity of a dynamical system to small perturbations in
its initial conditions over a finite time horizon \citep{Haller2015,
Shadden2005}. Given a flow map $\mathbf{F}_{t_0}^{t_0+T}$ that advances an
initial condition $\mathbf{x}_0$ at time $t_0$ to its state
$\mathbf{x}(t_0+T)=\mathbf{F}_{t_0}^{t_0+T}(\mathbf{x}_0)$ after an
integration time $T$, the sensitivity of the flow to a perturbation
$\delta\mathbf{x}_0$ is captured by the flow-map Jacobian
$\mathbf{J}=\partial \mathbf{F}_{t_0}^{t_0+T}/\partial \mathbf{x}_0$. The
right Cauchy--Green deformation tensor
\begin{equation}
\mathbf{C} = \mathbf{J}^{\mathsf T}\mathbf{J}
\label{eq:cauchy_green}
\end{equation}
is symmetric positive semi-definite, and its largest eigenvalue
$\lambda_{\max}(\mathbf{C})$ gives the maximum possible stretching of an
infinitesimal perturbation under the flow, attained for perturbations
aligned with the corresponding eigenvector. The FTLE is then defined as
\begin{equation}
\sigma_{t_0}^{T}(\mathbf{x}_0) = \frac{1}{|T|}\ln\sqrt{\lambda_{\max}(\mathbf{C})} ,
\label{eq:ftle}
\end{equation}
which gives the average exponential rate of separation, over the interval
$T$, of trajectories initialized infinitesimally close to $\mathbf{x}_0$.
Large values of $\sigma_{t_0}^{T}$ indicate initial conditions for which
nearby trajectories diverge rapidly -- i.e., directions in which the system
is highly sensitive to the initial state. Ridges of high FTLE are widely
used to identify the influence of stable and unstable invariant manifolds
and the associated transport barriers in cislunar dynamical systems
\citep{Haller2015}.

In this work, the FTLE is evaluated not in physical phase space, but over
the two-dimensional space of post-breakup ejection velocity perturbations
$(\delta v_x,\delta v_y)$ relative to the parent orbital velocity at the
breakup epoch, with the breakup position and the third velocity component
$\delta v_z=0$ held fixed. For each breakup phase, a regular
$(\delta v_x,\delta v_y)$ grid is propagated for $T=30$ days under the
BCR4BP, and the flow-map Jacobian is
approximated by central finite differences of the final position
$\mathbf{r}(T)$ with respect to $\delta v_x$ and $\delta v_y$:
\begin{equation}
J_{1} = \frac{\partial \mathbf{r}(T)}{\partial \delta v_x}, \qquad
J_{2} = \frac{\partial \mathbf{r}(T)}{\partial \delta v_y}.
\end{equation}
The corresponding $2\times2$ Cauchy--Green tensor is
$\mathbf{C}=[J_1\ J_2]^{\mathsf T}[J_1\ J_2]$, and
$\sigma_{t_0}^{T}(\delta v_x,\delta v_y)$ is computed from
Eq.~(\ref{eq:ftle}). In this velocity-space formulation, a high FTLE value
at $(\delta v_x,\delta v_y)$ indicates that fragments ejected with nearly
identical velocity perturbations in that neighborhood separate into
markedly different final positions after 30 days -- and, in particular, may
end up on opposite sides of the escape/retention boundary discussed in
Section~\ref{subsection:Escape_Mechanism}.

\section{Fragmentation Model}

The NASA Standard Breakup Model (SBM) \citep{2001AdSpR..28.1377J}, specifically the EVOLVE 4.0 version, is the most widely adopted framework for analyzing fragmentation events in the space environment. It serves as a foundational tool for short-term population environment predictions by determining the critical physical and dynamical characteristics of a resulting debris cloud, including characteristic size ($L_c$), area-to-mass ratio ($A/M$), and ejection velocity ($\Delta V$). These parameters are derived from statistical distributions estimated through empirical data gathered from documented on-orbit explosions, collisions, and extensive ground-based experimental databases.

It should be noted that the SBM was originally calibrated against fragmentation and collision events occurring in the LEO environment, and its empirical distributions for $L_c$, $A/M$, and $\Delta V$ do not explicitly account for cislunar space. In the absence of a breakup model specifically validated for cislunar fragmentation events, we adopt the EVOLVE 4.0 distributions as a representative, first-order characterization of the debris cloud, consistent with their use in other recent cislunar debris studies \citep{boone2021cislunar, Nishiguchi_etal_2025}. The resulting $\Delta V$ and $L_c$ distributions 
are regarded as a plausible benchmark scenario, and the propagation results presented below depend on BCR4BP.

\begin{figure}
  \centering
    \includegraphics[width=0.9\linewidth]{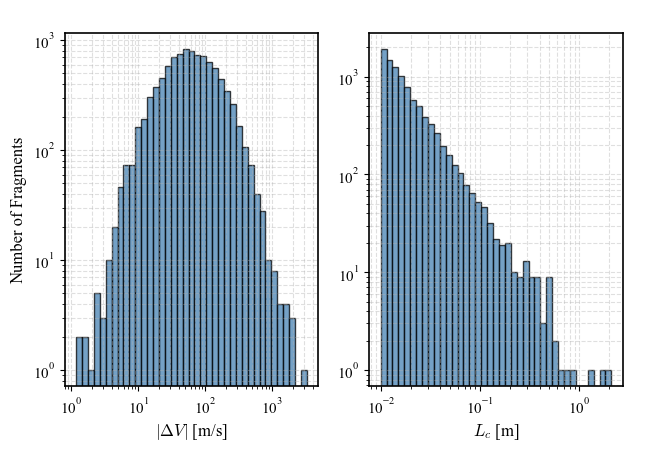}
    \caption{Distributions of the Magnitude of Ejection Velocity$|\Delta V|$(left), and Characteristic Length $L_c$(right)}
    \label{fig:num_distribution}
 \end{figure}
 
 \begin{figure}
    \centering
    \includegraphics[width=0.8\linewidth]{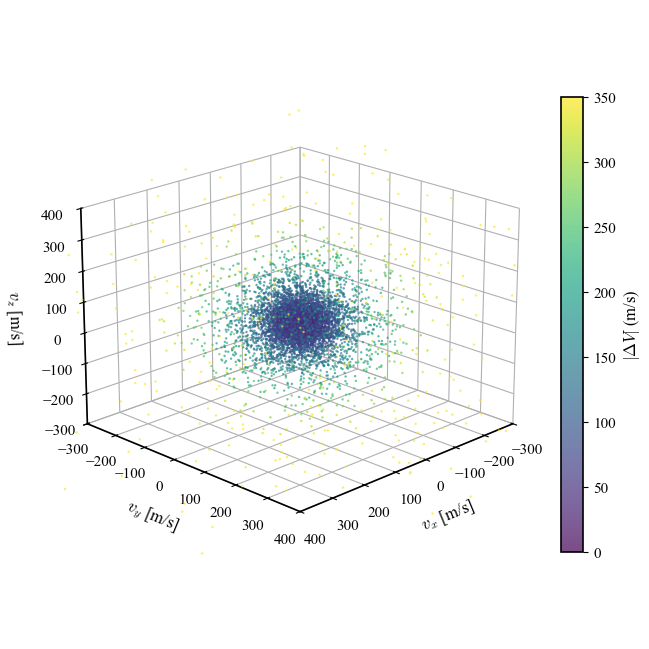}
\caption{Simulated Distribution of Breakup Ejection Velocities}
    \label{fig:velocitydistribution}
\end{figure}
SBM characterize debris properties via probabilistic distributions rather than deterministic states, as commonly adopted in debris environment modeling studies \citep{inproceedingsLetizia}. Sampling from these distributions enables the generation of a representative debris cloud for high-fidelity propagation using Monte-Carlo methods. Alternatively, these distributions can define the initial state uncertainty associated with a single fragment. 
 
\begin{table*}
  \centering
  \caption{Data of the analyzed DROs in the CR3BP model.}
  \label{tab:orbit_parameters}
  \begin{tabular}{lccccc} 
    \hline
    Orbit Name & Jacobi Constant & Period [nd]& Period [days] & Lunar perilune [km]& Example Missions \\
    \hline
    Orbit 1 & 3.0030 & 1.5342 & 6.8 & $\sim$37560  & / \\
    Orbit 2 & 2.9294 & 3.1767 & 14.1 & $\sim$68610  & Artemis I \\
    Orbit 3 & 2.7827 & 5.7174 & 25.3 & $\sim$157040  & DRO-A \\
    \hline
  \end{tabular}
\end{table*} 

In this study, we used the NASA Standard Breakup Model (EVOLVE 4.0) to simulate the fragmentation of a 1,000 kg spacecraft. The same event parameters are applied across all simulations to ensure the results are comparable. The specific data for the event are listed in Table \ref{tab:fragmentation_data}. To limit the computational load of the fragment propagation, the minimum size of the generated fragments is restricted to 0.01 m. A total of 9,409 fragments were generated by the model. Fig.~\ref{fig:num_distribution} displays the corresponding absolute frequency distributions of $\Delta V$, and $L_c$ of fragments generated from a breakup in a Distant Retrograde Orbit (DRO) at the orbital phase where the distance to Earth is minimal ($\phi = 0$), while Fig. \ref{fig:velocitydistribution} provides the illustration of the simulated ejection velocity.

\section{Dynamical Evolution of the Fragments}

\subsection{Orbit Selection}

\begin{figure}
    \centering
    \includegraphics[width=0.9\linewidth]{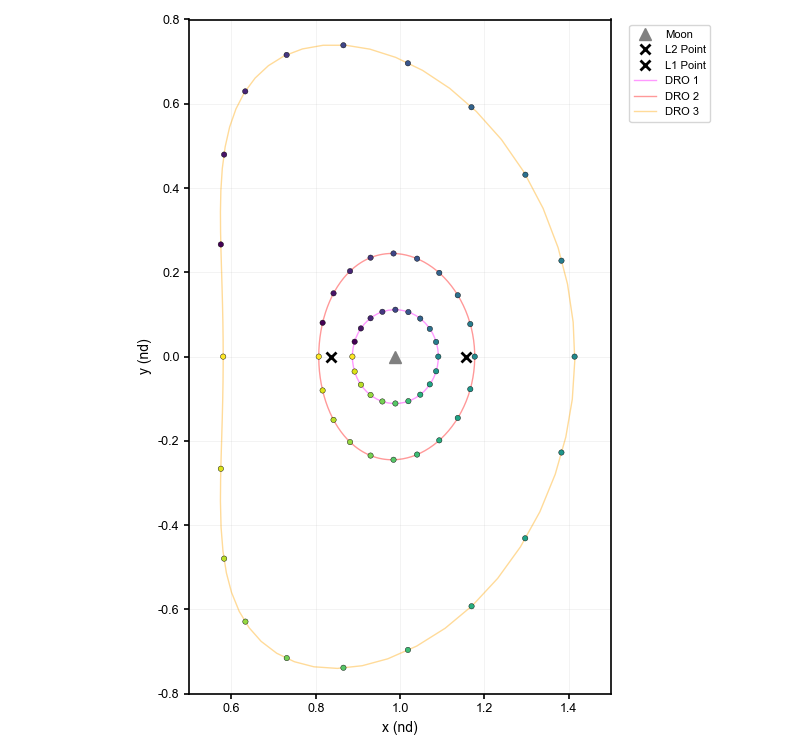}
    
    \caption{Three Distant Retrograde Orbits are modeled within the Earth-Moon CR3BP rotating frame. For each orbit, 20 breakup locations are sampled at equal time intervals to represent distinct orbital phases; the color-coding of these phases is consistent with the results presented in the subsequent sections.}
    \label{fig:dros}
\end{figure}

Distant Retrograde Orbits, in particular, are considered "natural habitats" \citep{inproceedingsSmitherman} within the cislunar domain due to their exceptional stability and low maintenance requirements. These orbits offer efficient transit corridors between the Earth, the Moon, and deep space, making them ideal candidates for hosting future lunar infrastructure. DROs constitute a unique family of periodic orbits within CR3BP, characterized by their clockwise (retrograde) motion around the Moon in the Earth-Moon rotating frame. The specific DRO trajectories analyzed in this study—designated as Orbit 1, Orbit 2, and Orbit 3—were generated using differential correction and numerical continuation techniques to sample different periods within the DRO family. These selected orbits are illustrated in Figure \ref{fig:dros}, and their data are summarized in Table \ref{tab:orbit_parameters}. For each orbit, 20 points are sampled at equal time intervals to accurately represent the debris evolution across different orbital phases.

A characteristic of these orbits is their resonance ratio, representing the ratio of the Moon's sidereal orbital period to that of the satellite. Notably, Orbit 2 is characterized by a 2:1 resonance, resulting in an orbital period of approximately 14 days and a periselenium altitude of roughly 68,000 km \citep{YHXB202510018}.

 \begin{figure*}   
  \centering  

    \includegraphics[width=\linewidth]{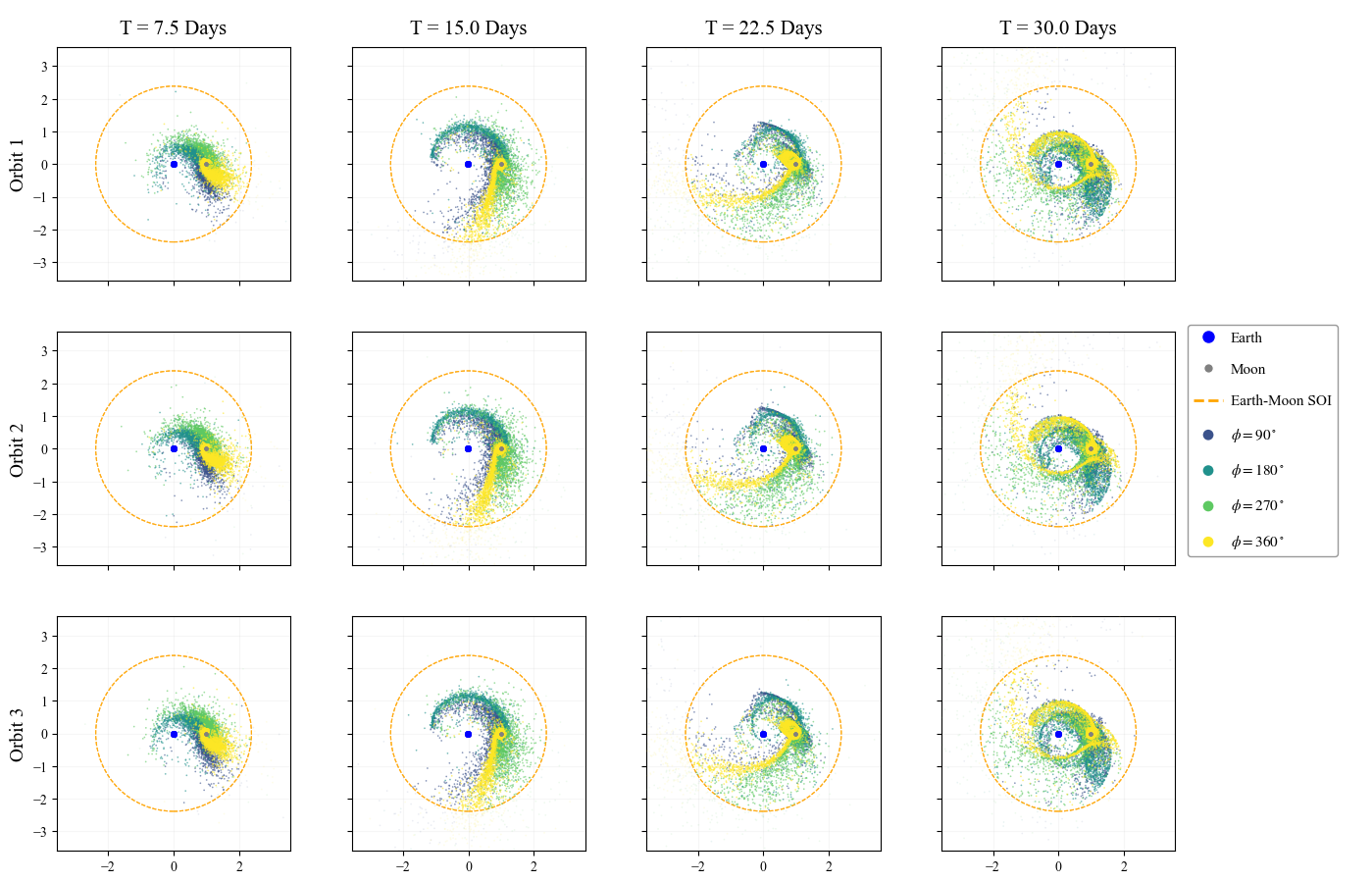}
    
    \label{fig:orbit1_cum}

\caption{Time-evolution of debris clouds originating from four distinct breakup points along Orbit 1, Orbit 2, and Orbit 3 trajectories. Each color represents a specific orbital phase: $\phi = \pi/2$ (indigo), $\phi = \pi$ (teal), $\phi = 3\pi/2$ (light green), and $\phi = 2\pi$  (yellow). Snapshots are taken at four progressive epochs within 30 days post-breakup in the Earth-Moon system.}
  \label{fig:droscatter}
\end{figure*}

\subsection{Spatial Dispersion}
\label{sec:dispersion}
This study employs the NASA Standard Breakup Model (SBM) to simulate the initial disintegration  of spacecrafts

, which are subsequently propagated under the BCR4BP (Eqs.~\ref{eq:bcr4bp_x}--\ref{eq:bcr4bp_z}) for the analyses presented in this and the following sections. 

The SBM provides a realistic distribution of energy and velocity, reflecting the complex spread characteristic of actual breakup events. To statistically sample varying conditions, multiple realizations are simulated by adjusting the initial orbital phase of the parent spacecraft. Twenty equally spaced points are selected along each reference orbit, starting from point 1 ($\phi = \pi/10$), with each independent fragment set propagated for a 30-day duration. Trajectories are terminated upon lunar impact (radius of 1738 km) or atmospheric reentry (altitude within 120 km of Earth’s surface).

A total of four breakup locations with equal orbital phase differences were selected for analysis. The evolution of the resulting fragment distributions across four time intervals (Fig.~\ref{fig:droscatter}) reveals distinct patterns that are highly sensitive to both the specific orbit and the initial breakup location.

Debris from Orbit 1 exhibits minimal dispersion, adhering closely to the progenitor’s path. In contrast, the debris from Orbit 3 exhibits the most extensive spatial dispersion among the three investigated cases. Specifically, fragments originating from point 20 ($\phi = 0$) demonstrate the widest scattering range—gradually expand to occupy a relatively vast region of cislunar space. While the expansion of fragments on orbit 3 is notably more pronounced than orbit 1 and orbit 2 scenarios, the clouds across all simulations retain a relatively cohesive structure. 

Despite these orbital-specific variations, a shared characteristic is the formation of debris "arms" (see Fig.~\ref{fig:drotraj}) that follow the dynamical flow of the Earth-Moon system. While dispersion scales vary between breakup locations, a consistent trend emerges: breakup events near Earth ($\phi = 0$) generally result in wider relative propagation and faster departure from the immediate lunar vicinity. This post-fragmentation dispersion is a highly non-uniform process, where the local state vector and the specific gravitational environment at the epoch of fragmentation exert a deterministic influence on the short-term evolution of the debris population. This sensitivity suggests that breakup events on the Earth-facing side or within larger-amplitude DROs result in wider relative propagation, presenting a heightened operational concern for mission safety within these stable regions. More details will be discussed in Section \ref{subsection:Escape_Mechanism}.

\begin{figure}
  \centering
    \includegraphics[width=0.9\linewidth]{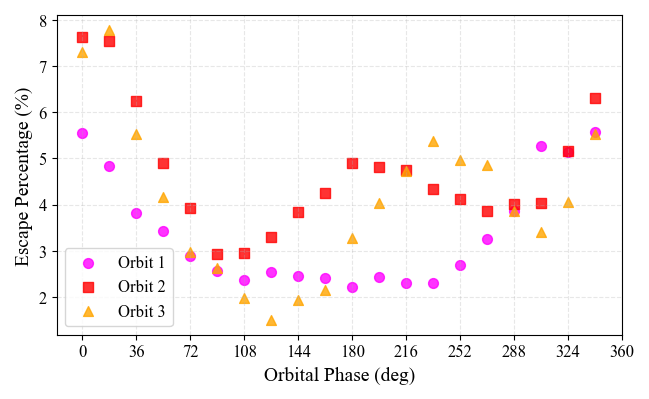}
  \caption{Escape Percentages Across Different Orbital Phases for Three Orbits (after 30-Day Propagation)}
  \label{fig:escapescatter}
\end{figure}

\subsection{Escape Mechanism}
\label{subsection:Escape_Mechanism}
In the context of this study, "escaped" is defined as the process by which fragments cross the Sphere of Influence (SOI) of the Earth-Moon system and traveling outside of it. The following analysis is divided into two primary drivers: the influence of the initial breakup location along the orbit and the magnitude and orientation of the resulting ejection velocities.
 
\subsubsection{Breakup Location}
The dynamical evolution of debris in DROs is affected by the initial breakup location. As illustrated in Figure \ref{fig:escapescatter}, the escape percentage exhibits a pronounced phase-dependent behavior across all three surveyed orbits, exhibiting a periodic trend. Specifically, breakups occurring on the Earth-facing side ($\phi \approx 0^\circ \text{ or } 18^\circ$) yield the highest escape risk, with peaks reaching approximately $8\%$. In contrast, the stability increases substantially when the breakup occurs on the lunar far-side ($100^\circ < \phi < 200^\circ$), where escape rates for all orbits drop below $3\%$, and Orbit 3 reaches its minimum of less than $1.5\%$ at $\phi = 126^\circ$. Among the cases studied, the fragments breakup 
originated from orbit 1 remain the most stable on a 30-day duration, maintaining a consistently lower escape fraction across most orbital phases compared to Orbits 2 and 3.
\begin{figure*}
  \centering
    \includegraphics[width=0.95\linewidth]{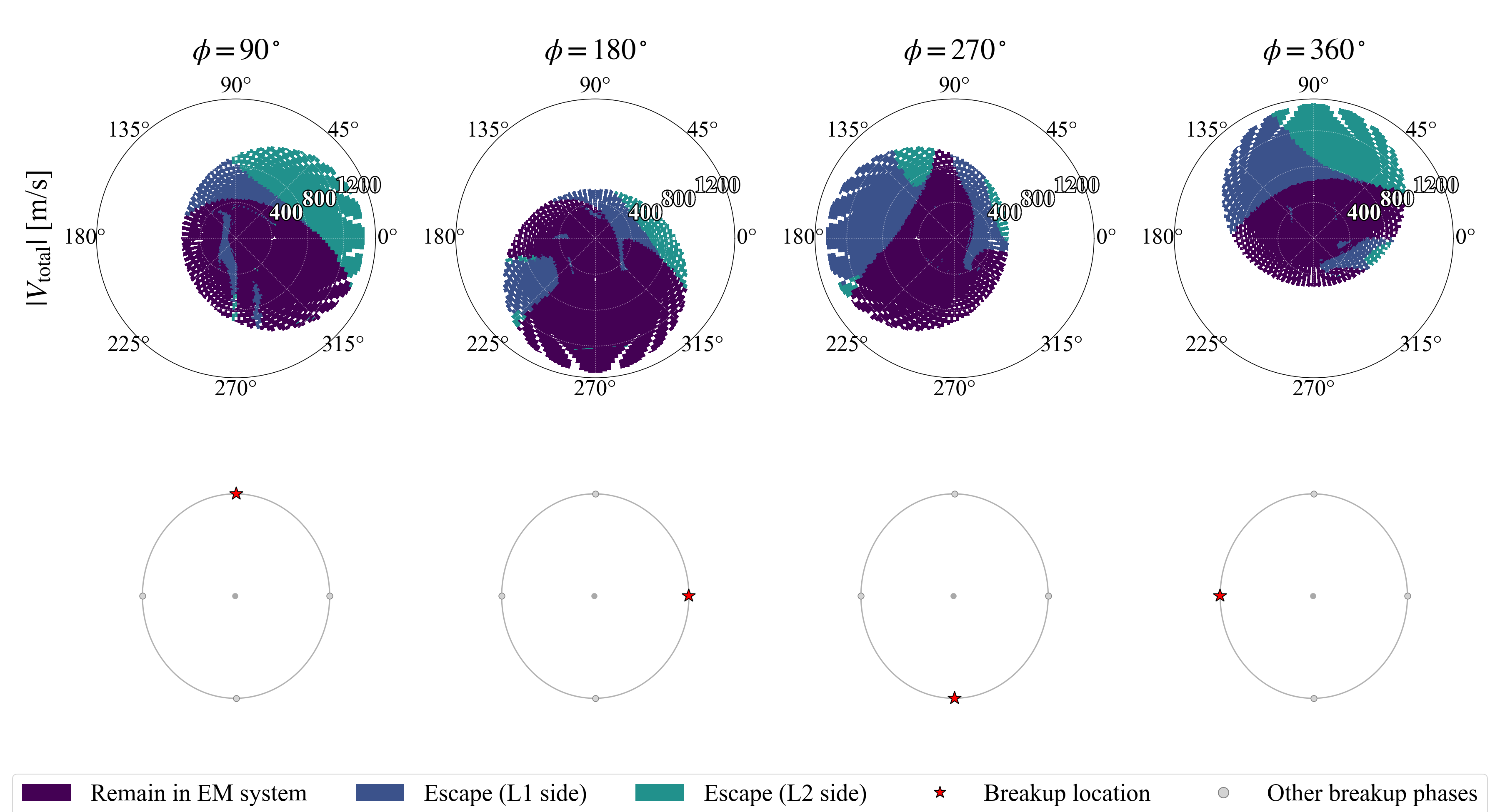}
  \caption{Fate of fragments on Orbit 1 with different velocity increments shown in polar coordinates $(\theta_{total},|V_{total}|)$, where $\theta_{total}$ and $|V_{total}|$ are the direction and magnitude of the total post-breakup velocity in the rotating frame. Each point is colored by its 30-day fate. Uncovered cells are left blank. The bottom row indicates the corresponding breakup location (red star) along the reference orbit.}
  \label{fig:velocities}
\end{figure*}
\begin{figure*}
  \centering
    \includegraphics[width=0.95\linewidth]{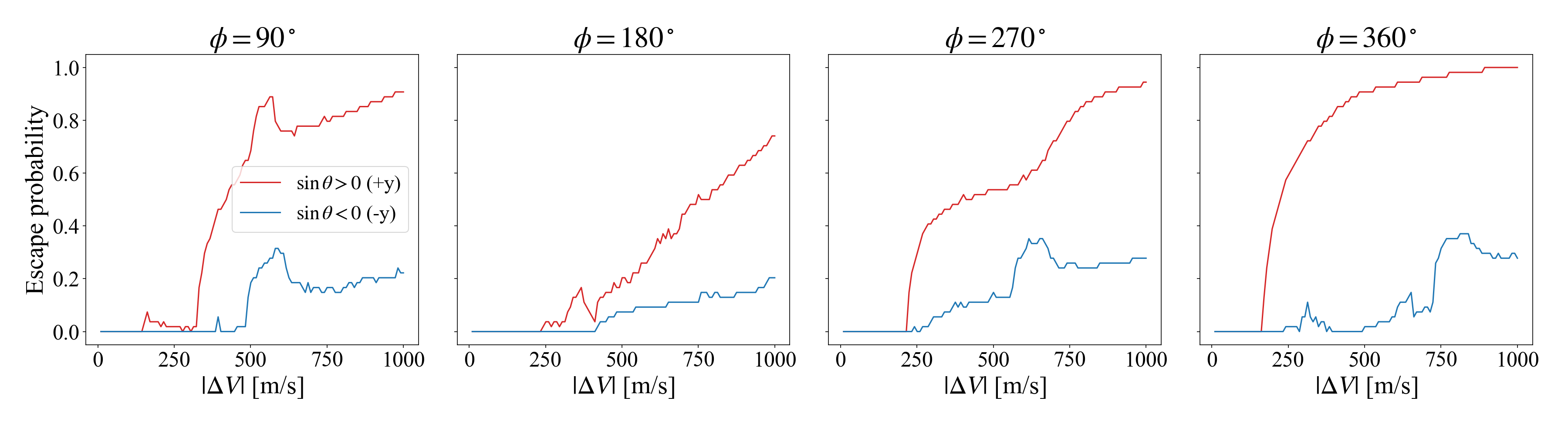}
  \caption{Escape probability as a function of $|\Delta V|$ for ejections into the $+y$ half-plane (red) versus the $-y$ half-plane (blue), for each of the four breakup phases.}
  \label{fig:H4_coriolis}
\end{figure*}
\begin{figure*}
  \centering
    \includegraphics[width=0.95\linewidth]{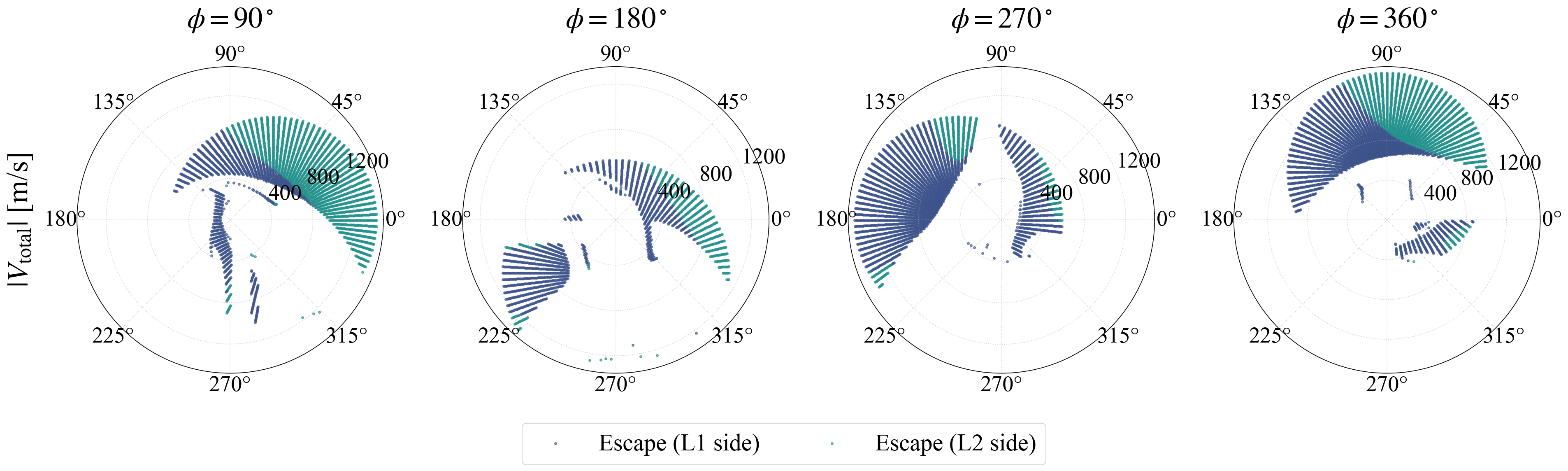}
  \caption{Polar distribution of the total post-breakup velocity $(\theta_{total},|V_{total}|)$ for fragments that escape on the $L_1$ side (blue) versus the $L_2$ side (teal), for each of the four breakup phases.}
  \label{fig:vtotal_l1l2}
\end{figure*}
 \begin{table}[h]
\centering
\caption{Orbit 1: States at different phases}
\label{tab:state_vectors}
\begin{tabular}{ccc}
\hline
Phase ($\phi$) & Position $[x, y, z]$ [nd] & Velocity $[v_x, v_y, v_z]$ [nd] \\ \hline
$90^\circ$  & [0.9888, 0.1114, 0.0]&[0.4084, -0.0025, 0.0]  \\
$180^\circ$ & [1.091, -0.0006, 0.0]&[-0.0022, -0.4622, 0.0]  \\
$270^\circ$ & [0.9881, -0.1114, 0.0]&[-0.4083, 0.0006, 0.0] \\
$360^\circ$ & [0.887, 0.0, 0.0]&[0.0, 0.4711, 0.0]  \\ \hline
\end{tabular}
\end{table}

\subsubsection{Velocity Magnitude Thresholds}
Figure \ref{fig:velocities} illustrates how the escape percentage relates to the magnitude and direction of the ejection velocity as well as the total post-breakup velocity. These data points are derived from a uniform spherical sampling of velocities ranging from 3.43 to 983.70 m/s, which effectively encompasses 99.7\% of the probable ejection velocity distribution defined by the NASA Standard Breakup Model (SBM) of Orbit 1.

The escape risk is highly anisotropic. For impulses with $|\Delta V| < 200$ m/s, escape probability remains near zero regardless of the ejection angle, indicating a minimum energy threshold for fragments to exit the Earth-Moon system within a 30-day window. At higher speeds ($>600$ m/s), fragments escape across a much broader range of velocity orientations.

\subsubsection{Directional Preference}
As shown in Figure \ref{fig:velocities}, at various orbital phases, velocity increments that result in a total post-breakup velocity directed between $0^\circ$ and $180^\circ$ are the most conducive to escape.  Specifically, for the $\phi = 360^\circ$ case, a substantial portion of fragments with velocity angles between $45^\circ$ and $135^\circ$ depart the Earth-Moon SOI. For all investigated phases, fragments ejected toward the $+y$ direction ($\theta = 90^\circ$) reach an escape rate approaching $100\%$ once the ejection magnitude exceeds $600$ m/s.

The structures visible in Figure \ref{fig:velocities} likely reflect a combination of (i) the post-breakup Jacobi constant, which decreases as $|\Delta V|$ grows and can open the $L_1$/$L_2$ necks once it drops below the corresponding critical values, allowing fragments to begin escaping; and (ii) the magnitude and the orientation of the fragment's velocity, which is set by the ejection geometry and differs with the orbital phase $\phi$. The first effect is broadly consistent with the growth of the escaped region with $V_{total}$.

The asymmetry in escape probability is quantified in Figure \ref{fig:H4_coriolis}. Across all breakup phases, ejections with a $+y$ velocity component escape 3.4 to 6.6 times more often than those with a $-y$ component. This likely results from the Coriolis acceleration $-2\boldsymbol{\Omega}\times\mathbf{v}$, which deflects $+y$ and $-y$ ejections asymmetrically in the rotating frame. Since this effect depends only on the sign of the $y$-velocity, it persists regardless of the breakup phase.

The fragments escaping on the \(L_1\) side (Earth side) and those escaping on the \(L_2\) side (outer side) have mean velocity directions of approximately \(206^\circ\) and \(26^\circ\), respectively (see Figure~\ref{fig:vtotal_l1l2}). This corroborates the intuitive expectation that the escape gateway (\(L_1\) or \(L_2\)) is strongly correlated with the direction of the total post-breakup velocity vector, i.e., velocities pointing toward \(L_1\) preferentially exit through \(L_1\), and similarly for \(L_2\).



\subsubsection{Phase-Space Geometry of the "Escape Boundary"}

The boundary is defined as a sampled point (cell) for which at least one neighboring point has a different escape outcome (escape vs. retention), it marks the transition between qualitatively different fates. The polar maps in Figure~\ref{fig:velocities} show that the escape and
retention regions are not separated by a smooth curve, but by a boundary
with a special structure. To understand the origin of this
structure, we compute the finite-time Lyapunov exponent (FTLE) over a
Cartesian ejection-velocity space $(\delta v_x,\delta v_y)$ centered on the
parent velocity, restricted to $\sqrt{\delta v_x^2+\delta v_y^2}\le 983.70$
m/s. Large FTLE values indicate directions in velocity
space along which nearby trajectories diverge rapidly -- i.e., ridges of
strong sensitivity to initial conditions.

\begin{figure*}
  \centering
    \includegraphics[width=0.95\linewidth]{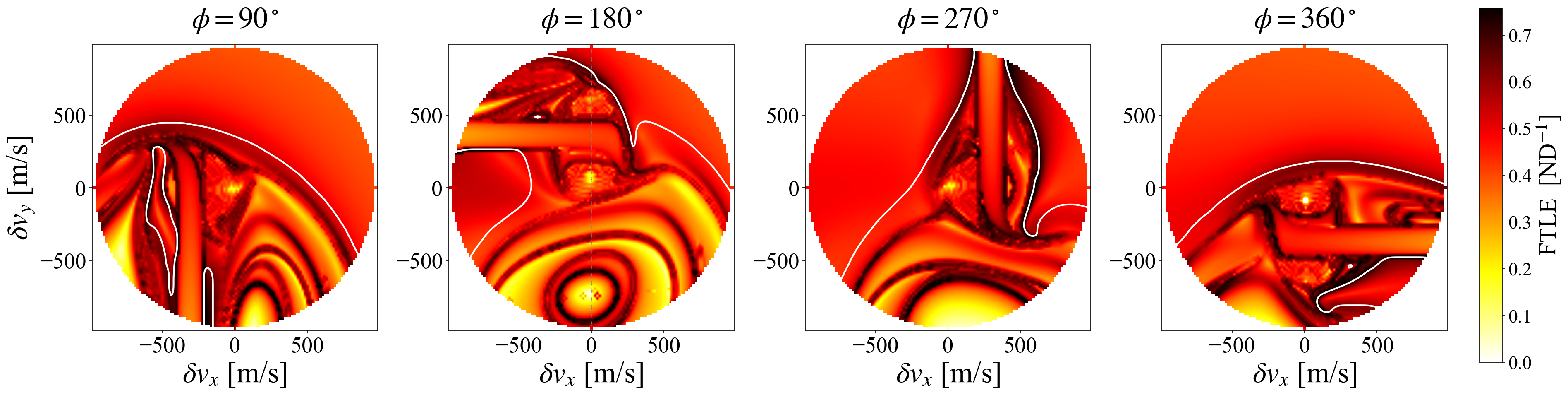}
  \caption{{Finite-time Lyapunov exponent (FTLE) in ejection-velocity space
  $(\delta v_x,\delta v_y)$ for the four breakup phases. The white curve
  marks the 50\% escape-probability contour, i.e., the escape/retention
  boundary.}}
  \label{fig:ftle}
\end{figure*}

As shown in Figure~\ref{fig:ftle}, the escape/retention boundary (white
curve) runs along the ridges of high FTLE in every phase. This
correspondence is also borne out quantitatively: cells on the boundary are
4.9 to 7.6 times (across the four phases) more likely
than a randomly chosen cell to fall within the top 5\% of the FTLE
distribution. This indicates that the boundary is a sensitive dividing line
in the sense of chaotic scattering: ejection conditions on either side of
it are dynamically similar, but infinitesimally small differences in
$|\Delta V|$ or direction are amplified over the 30-day propagation into
qualitatively different fates (escape or stay in the earth-moon SOI). The pattern of the
boundary in Figure~\ref{fig:velocities} is therefore a direct consequence of
this underlying nonlinear (chaotic) dynamics, rather than of a simple
energy-threshold (Jacobi constant) criterion. A more detailed discussion of this phenomenon is left for future work.

\section{Environmental Impact Assessment}
\label{sec:env}
\subsection{Moon Impact}
Lunar contamination may arise from orbital debris impacting the Moon. To comply with planetary protection and environmental guidelines \citep{nasa2020planetary}, lunar missions are required to undergo comprehensive risk assessments. Here, we analyze the escape and impact percentages of fragments resulting from a breakup on Orbit 1 (DRO); a detailed discussion of impact dynamics and site-specific consequences will be discussed in future work.

\begin{figure}
  \centering
    \includegraphics[width=0.9\linewidth]{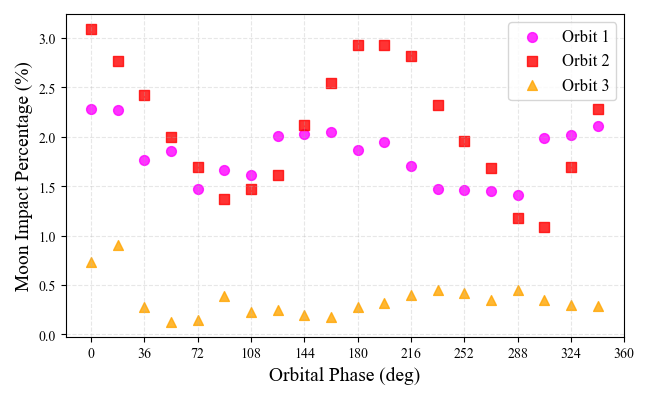}
  \caption{Moon Impact Percentages Across Different Orbital Phases for Three Orbits (after 30-Day Propagation)}
  \label{fig:impactscatter}
\end{figure}

The simulation results (Figure \ref{fig:impactscatter}) indicate that the cumulative percentage of fragments striking the lunar surface remains low, staying below 3.5\% for all cases over the 30-day propagation duration. Among the three configurations, orbit 2 exhibits the highest relative impact rate and the most significant fluctuations, with its impact percentage showing periodic oscillations that peak at $\phi = 0$ (approximately 3.1\%) and $\phi = 180^\circ$ (approximately 2.9\%), while dropping to a minimum of 1.1\% at $\phi = 306^\circ$ . In contrast, orbit 1 demonstrates a more stable risk profile, with impact percentages fluctuating moderately between 1.4\% and 2.3\% across all breakup locations. Orbit 3, owing to its greater proximity to the Earth-Moon system's outer regions, maintains an impact risk consistently below 1.0\% across all phases, often hovering between 0.1\% and 0.5\%. Overall, breakups occurring near the lunar proximal side (phases) tend to yield a higher initial impact risk for both orbit 1 and orbit 2. The evolution of these impact rates over time for the various scenarios is further illustrated in Figure \ref{fig:moonimpact}.

\begin{figure}
  \centering
    \includegraphics[width=0.9\linewidth]{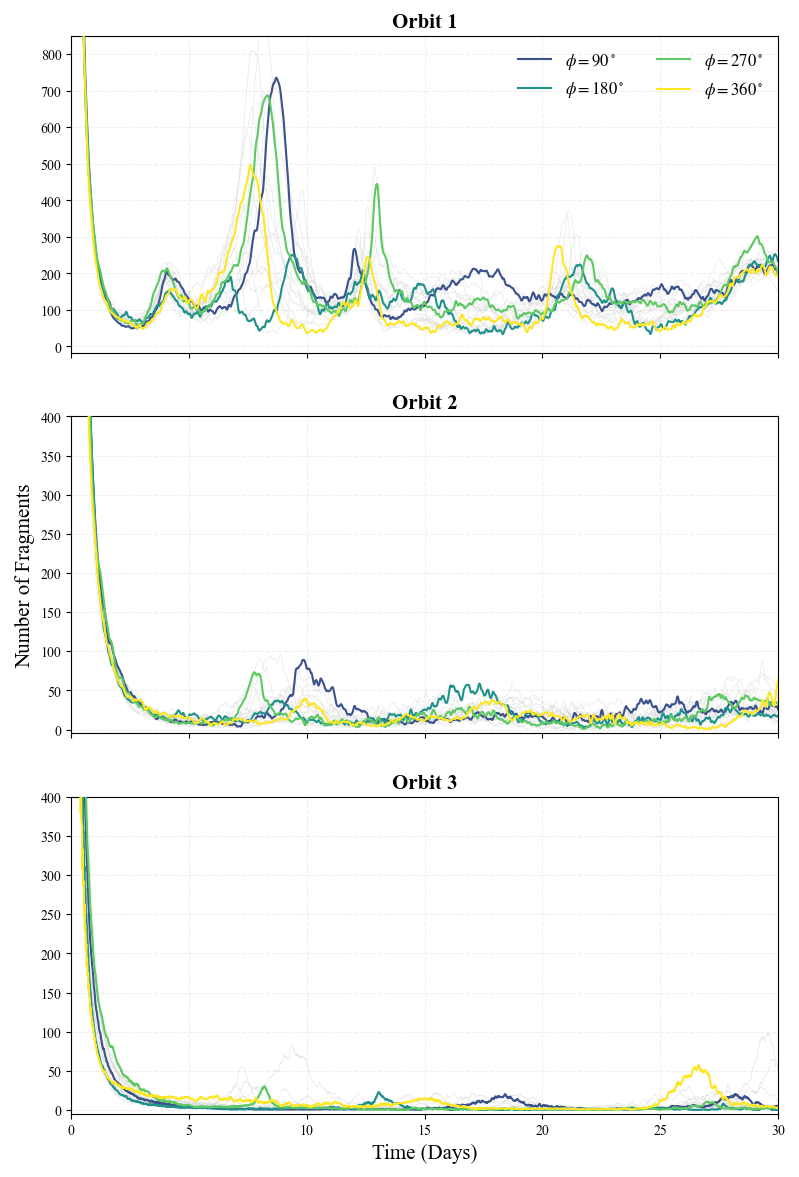}
  \caption{Percentage of fragments entering the vicinity of original orbit over a 30-day propagation. Representative orbital phases are highlighted in color, with all other phases shown in grey.}
  \label{fig:unsafe_fraction_comparison}
\end{figure}

\subsection{Orbital Interference}
Fragmentation events in periodic cislunar orbits produce artificial debris that threatens the orbital ecosystem. A recent study by \citet{boone2021cislunar} demonstrates that these fragments circulate through wide-ranging cislunar regimes, potentially intercepting other spacecraft near the site of the original mishap. Consequently, debris-generating events in these orbits necessitate a global assessment of cislunar space safety.

For fragmentation events occurring on a DRO, an assessment is adopted to provide an estimate of the potential interference between the resulting debris cloud and the parent spacecraft during the early post-breakup phase. Rather than explicitly modeling the spacecraft as a moving object—which would introduce an additional degree of freedom associated with its unknown orbital phase and thus reduce the generality of the analysis—the spacecraft’s potential locations are encapsulated by a static "protected region" surrounding the nominal DRO trajectory. This treatment eliminates the need to prescribe the spacecraft’s instantaneous phase along the orbit. In the present study, the protected region is defined as a toroidal shell enveloping the nominal DRO before breakup. Specifically, this volume is constructed as a tube with a radius of 200 km centered along the reference trajectory. This geometry effectively creates a "donut-shaped" protected zone that follows the entire path of the orbit. The 200 km radius is based on the fact that DROs possess exceptionally low sensitivity to state errors and perturbations. This stability ensures a slow natural drift without station-keeping; typical position deviations remain within 10-30 kilometers over 7 days \citep{AoHaiyue2024Stationkeeping}.  

\begin{figure}
  \centering
    \includegraphics[width=0.9\linewidth]{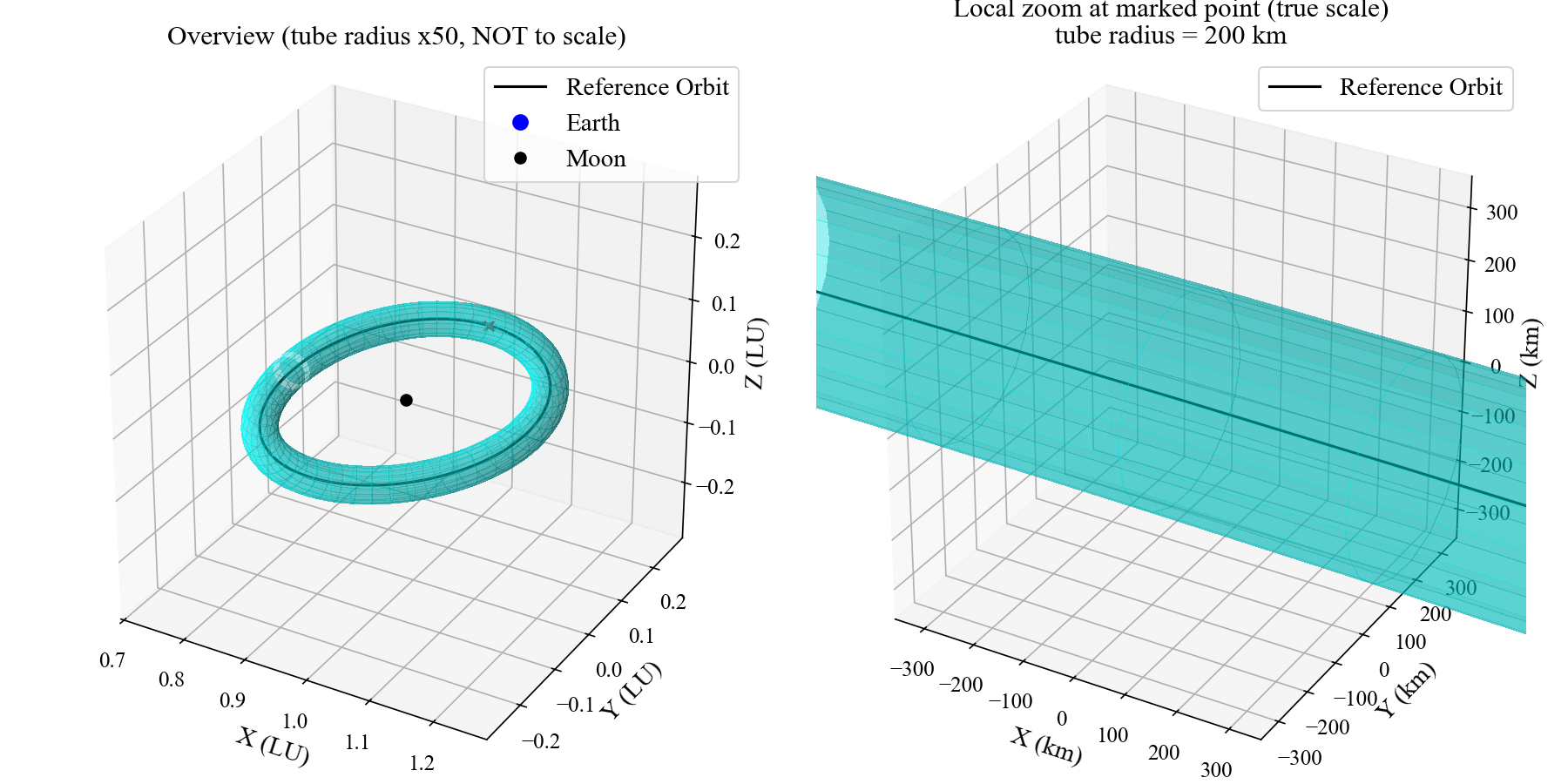}
  \caption{Illustration of the DRO vicinity region (defined as the "Protected Region") in the Earth-Moon CR3BP rotating frame.}
  \label{fig:droregion}
\end{figure}

\begin{figure}
  \centering
    \includegraphics[width=0.9\linewidth]{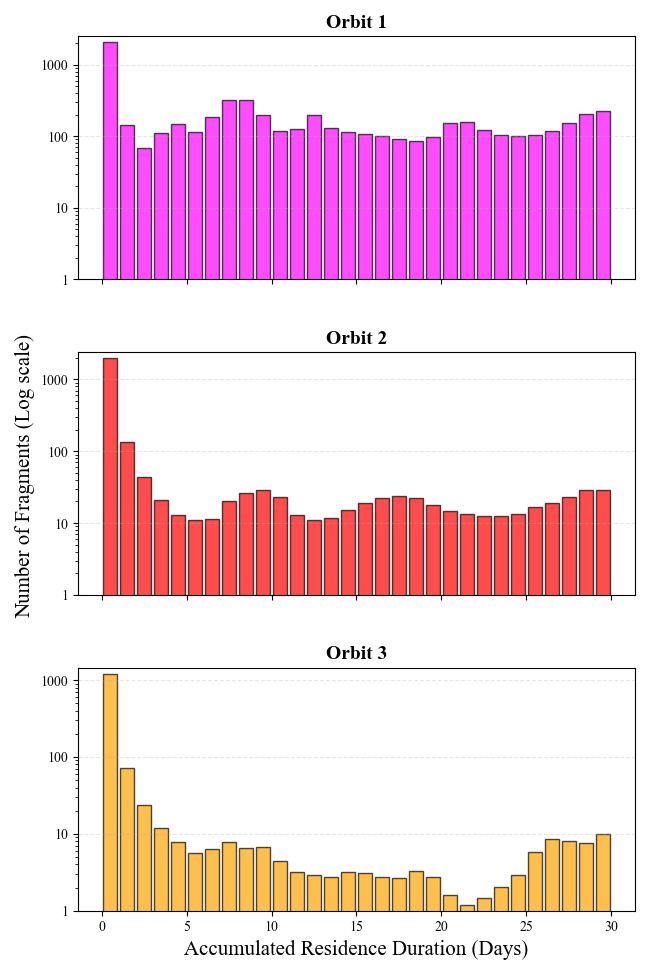}
  \caption{Accumulate residence duration of fragments entering the vicinity of original orbit over a 30-day propagation. }
  \label{fig:acctime_comparison}
\end{figure}

A debris fragment is considered to interfere with the spacecraft on the parent orbit if it enters this area after the fragmentation event. This definition provides a conservative and phase-independent criterion for early-stage interference assessment. The adopted protected region is illustrated in Fig. \ref{fig:droregion}. We assess debris incursions into "protected regions" \citep{IADC_2025_Mitigation_Guidelines} using two metrics: (i) the percentage of fragments that interfere with the protected zone, and (ii) the cumulative time fragments spend inside the zones. This approach accounts for both the uncertainty in the spacecraft's position and the variability in the initial locations of the fragmentation event. 

The initial samples are propagated in the BCR4BP model for 30 days for each orbital phase. The short-term evolution of debris clouds initiated from various DROs reveals distinct dispersion mechanisms. The results (see Fig.~\ref{fig:unsafe_fraction_comparison}) reveals distinct interference mechanisms under the 200 km safety zone. The overall entry rate generally remains below $3.0\%$, reflecting the effective dilution of debris clouds near the parent orbit.

The re-entry rate for orbit 1 exhibits an initial surge followed by a series of decaying oscillations, with a notable transient re-entry peak emerging around the seventh day. The cumulative residence time distribution (Fig.~\ref{fig:acctime_comparison}) shows that while most interference events are transient—with a primary peak at less than 1 day—the average residence time remains relatively high. This confirms that a higher proportion of fragments from an orbit 1 fragmentation event tends to linger in the vicinity of the parent orbit compared to other cases.

The fragmentation events on orbit 2 are characterized by periodic re-entry spikes, most notably around day 10 and day 28, where the entry rate briefly climbs to approximately $0.1\%$. This suggests that although orbit 2 exhibits a "self-cleaning" property in the short term, the underlying orbital mechanism compels fragments to periodically re-traverse the spacecraft's vicinity. The residence times are concentrated within very short durations (peak < 1 days), indicating that these encounters are primarily high-velocity, fly-by events rather than persistent co-orbital threats.

Fragmentation on orbit 3 displays the lowest re-entry risk profile among the three cases. Re-entry rates remain low, with only occasional events observed around day 7 and day 26. The residence time distribution in orbit 3 is the most restricted, with most fragments spending less than 2 days within the protected region over 30 days. This indicates that a fragmentation event in orbit 3 leads to rapid dispersion into the broader cislunar environment, posing the least threat to the parent spacecraft's immediate operational safety.

It is instructive to compare these results with prior assessments of debris generated on Halo orbits and NRHOs near the collinear libration points \citep{articleBlack,2023amos.conf...20B,boone2021cislunar,2023JSpRo..60..668B}. Those studies generally report that fragments released on Halo/NRHO orbits disperse away from the parent trajectory considerably faster, and depart the Earth--Moon SOI in substantially larger proportions, often within days rather than weeks. This contrast can be understood in terms of the underlying dynamical environment: Halo orbits and NRHOs reside close to the unstable $L_1$/$L_2$ collinear points, where the local manifold structure provides pathways for fragments to escape the Earth-Moon SOI in 30 days. By contrast, DROs occupy a more dynamically stable region around the Moon, as quantified in Sections~\ref{sec:dispersion}--\ref{subsection:Escape_Mechanism}. Consequently, escape fraction remain 
low over 30 days for most breakup phases. The majority of fragments are loosely bound and slowly dispersing within the Earth-Moon SOI (Fig.~\ref{fig:droscatter}). In this sense, the DRO debris environment is characterized by: (i) the long-lived presence of fragments near the Moon, which contributes to the lunar impact rates discussed above; and (ii) the periodic re-entries into the immediate vicinity of the parent orbit, as documented in Fig.~\ref{fig:unsafe_fraction_comparison} and Fig.~\ref{fig:acctime_comparison}. This suggests that, relative to Halo/NRHO debris, DRO fragmentation events pose a more persistent, localized risk to co-orbital assets and to the lunar surface.

\section{Conclusion}

This study provided a comprehensive dynamical analysis of spacecraft fragmentation in Distant Retrograde Orbits within the Earth--Moon system. By propagating debris clouds generated from the NASA Standard Breakup Model across varying orbital phases, we identified three distinct risk regimes governed by the inherent stability and resonant characteristics of the progenitor orbits.

A critical finding of this research lies in the fragment escape and retention patterns, which exhibit a profound dependence on the initial breakup phase. Our analysis reveals that debris generated in the lunar far-side region tends to remain trapped within the Earth--Moon sphere of influence (SOI) in larger proportions compared to Earth-side breakups. Orbit 1 and orbit 3, in particular, demonstrate high short-term retention rates exceeding 90\% after 30 days. In contrast, orbit 2 presents a special case where fragments released near the Earth-facing side depart the system rapidly, with retention rates falling below 10\%, while far-side breakups maintain a retention rate higher than 95\%. This disparity suggests that the dissipation of a debris cloud in cislunar space does not occur uniformly but is strictly governed by the local dynamical environment at the epoch of fragmentation. 

Furthermore, the statistical analysis of post-fragmentation states reveals a pronounced directional anisotropy in the debris escape mechanism. 

 The likelihood of a fragment departing the cislunar system is highly sensitive to the orientation of the velocity vector. Specifically, fragments ejected with velocity vectors oriented near $90^\circ$ (the $+y$ direction) exhibit the highest escape rate, frequently reaching near-unity escape probability even at moderate impulse magnitudes. This asymmetry is quantified across all breakup phases, where ejections with a $+y$ velocity component escape 3.4 to 6.6 times more often than those with a $-y$ component, a disparity attributed to the Coriolis acceleration in the rotating frame. Moreover, fragments escaping through the $L_1$ and $L_2$ escape gateways correlates strongly with the orientation of the post-breakup velocity vector. Beyond this directional preference, the boundary separating escape and retention exhibits a complex structure rather than a smooth energy threshold. FTLE analysis reveals that this boundary aligns with ridges of high Lyapunov exponent, where infinitesimally small changes in $|\Delta V|$ or ejection direction are amplified into qualitatively different fates over the 30-day propagation window. Predicting the short-term evolution of a debris cloud requires not only the spatial coordinates of the breakup but also a precise characterization of the fragmentation event's energy and location.

Regarding collision risks, the probability of direct lunar impact remains negligible across all orbits, with cumulative impact rates staying below $2.5\%$. Furthermore, the numerical results demonstrate that the risk posed by a fragmentation event to the parent orbit over 30 days is low. Under the 200 km safety criterion, instantaneous entry rates consistently remain below $3.0\%$, suggesting that the debris cloud undergoes rapid dilution in cislunar environment.

These findings underscore the necessity for phase-dependent risk assessment in cislunar mission planning. The data reveal that the re-entry of a breakup event on DROs is not only correlated with the orbital phase at the epoch of fragmentation but also the orbit size. Orbit 1 exhibits a relatively higher average residence time for fragments, while orbit 2 and orbit 3 benefit from "self-cleaning" mechanisms or rapid dispersion. However, even in these unstable cases, periodic re-entry events can create transient windows of collision risk. 

Future research should further investigate the underlying mechanisms that dictate the steering of debris out of the Earth-Moon SOI, as well as the drivers of periodic re-entry behaviors. A deeper understanding of how the combination of velocity magnitude and direction governs the dynamical fate of fragments will be essential to refining predictive avoidance maneuvers.

\section*{Acknowledgements}
This work was supported by the National Natural Science Foundation of China (Grant Nos. 12432017 and 12303082). 
Peng Shu also acknowledges the Yunnan Revitalization Talent Support Program.
\appendix

\section{Additional results}

\subsection{Temporal Evolution of Fragment Escape Rates}
\begin{figure} 
  \centering

    \includegraphics[width=\linewidth,height=0.72\textheight,keepaspectratio]{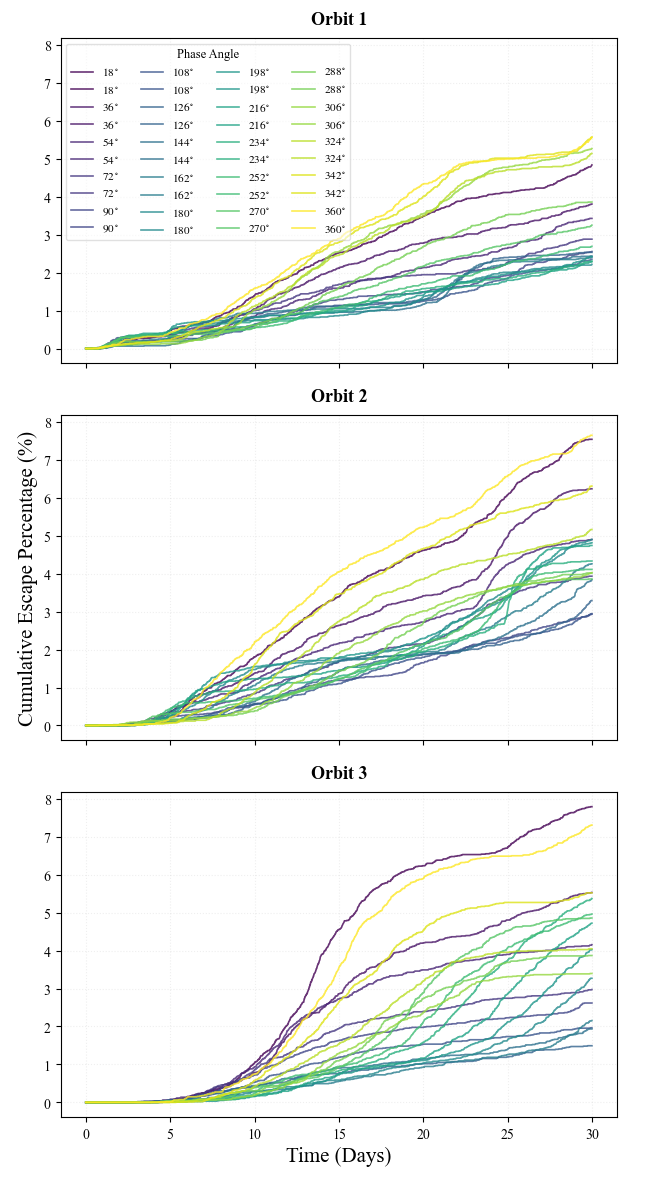}

\caption{Escape fraction of fragments on Orbit 1,  Orbit 2 and Orbit 3 in the Earth-Moon system for 30 days, initiated from four different points along the trajectory.}
  \label{fig:droescape}
\end{figure}

 Figure \ref{fig:droescape} presents the cumulative escape fraction of debris clouds over a 30-day propagation period for the three candidate orbits, with each individual curve corresponding to one of the 20 discrete breakup points along the trajectory. The data reveals that the escape process is characterized by a non-linear growth pattern, typically maintaining stability for the first 3 to 5 days before exhibiting a gradual increase in the percentage. Notably, the escape rate shows a strong dependence on the initial breakup phase, particularly for orbits 2 and 3, where certain points trigger rapid surges in the escape fraction between Day 10 and Day 20. By the end of the simulation, the total escape percentage stabilizes between 2\% and 8\%, reflecting the varying degrees of dynamical sensitivity to the initial conditions within the Earth-Moon system.

\subsection{Temporal Evolution of Lunar Impact Risk}
\begin{figure}
    \centering
    \includegraphics[width=\linewidth,height=0.72\textheight,keepaspectratio]{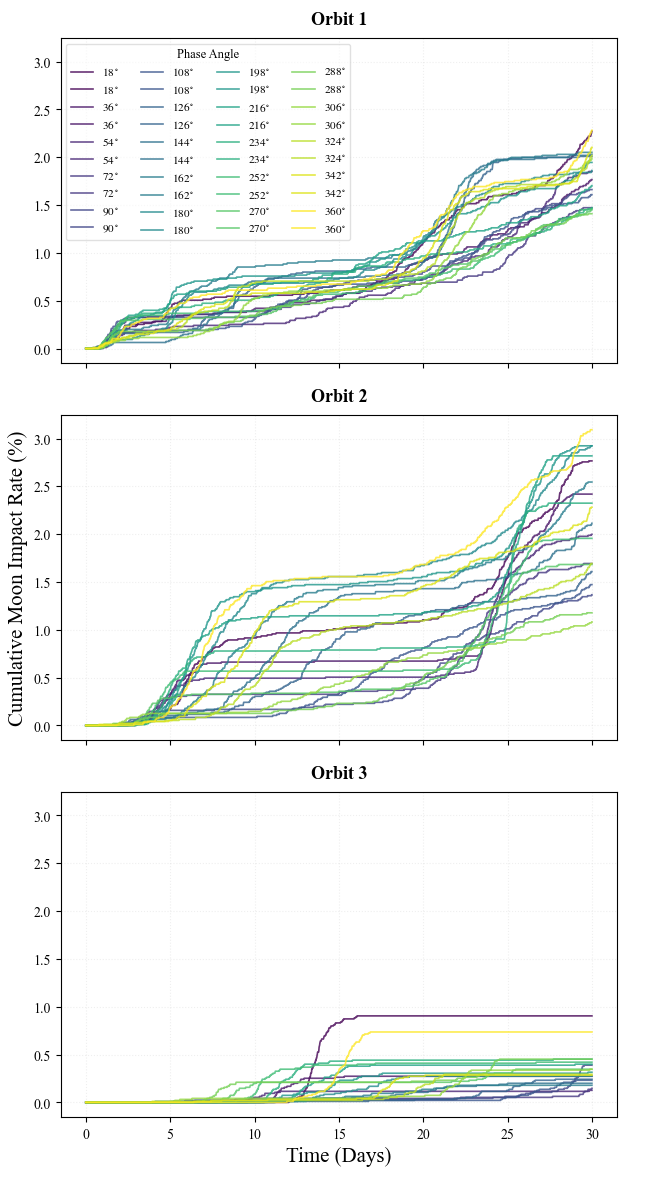}
 
    \caption{Cumulative percentage of fragments impacting the Moon 30 days after breakup events occurring at points 1-20 on orbit 1, orbit 2 and orbit 3.}
    \label{fig:moonimpact}
\end{figure}

The temporal distribution of lunar impact risk is illustrated in Figure \ref{fig:moonimpact}, showing the cumulative percentage of fragments impacting the Moon's surface. Unlike the relatively smooth growth seen in escape fractions, lunar impacts tend to occur in discrete, step-like increments. While orbit 1 and 2 exhibit a persistent and multi-stage increase in impact risk that often intensifies after Day 20, orbit 3 remains stable for the majority of the duration. For nearly all breakup points in orbit 3, the impact risk stays near zero for the first 10 days, followed only by a minor late-stage surge that plateaus well below 1.0\%. 

\subsection{Full Trajectories of Fragments}
 
\begin{figure*} 
  \centering
    \includegraphics[width=0.95\linewidth]{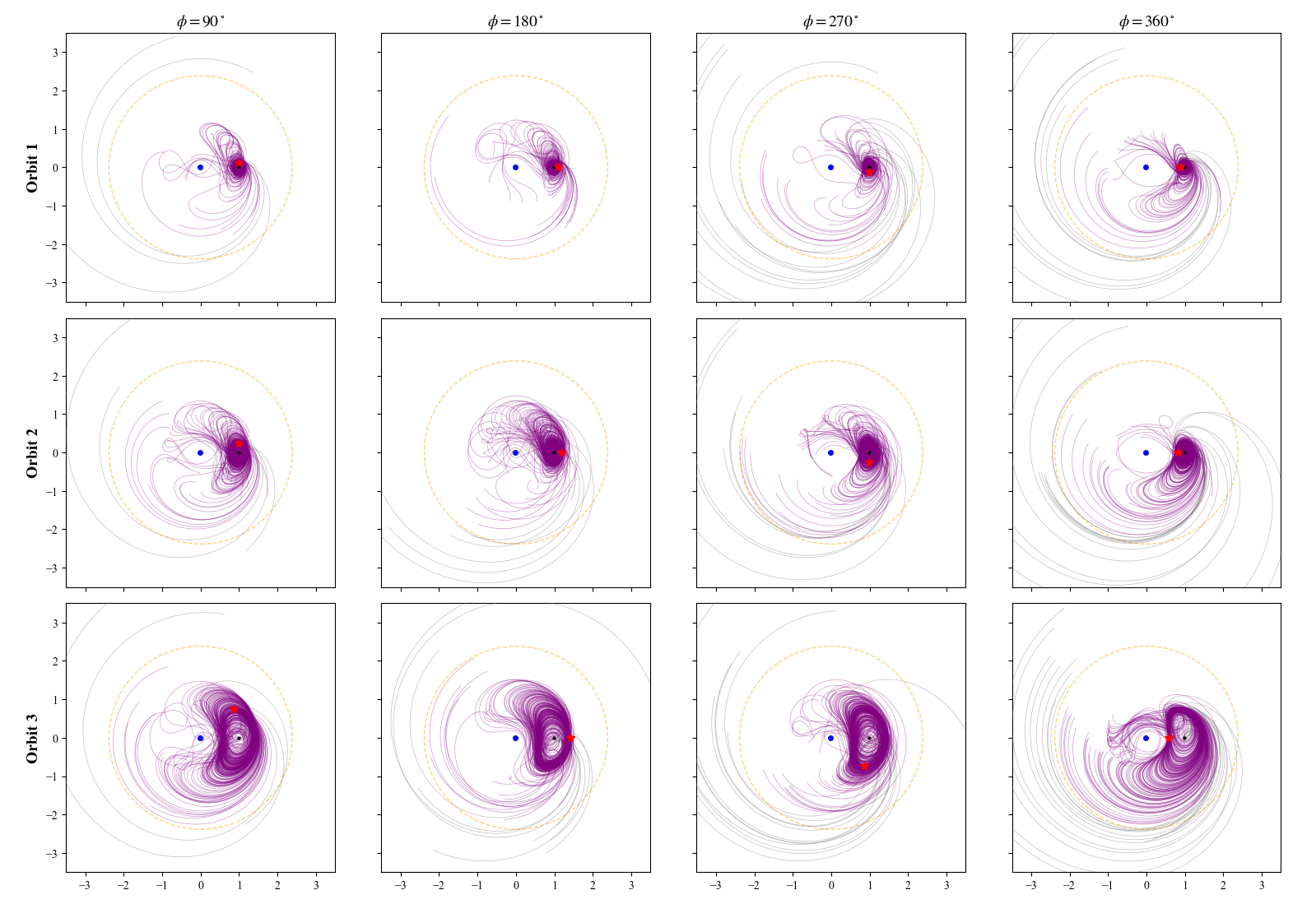}
\caption{Dynamical evolution of debris clouds on orbit 1, orbit 2 and orbit 3 in the Earth-Moon system for 30 days, initiated from four different points along the trajectory. Purple markers denote fragments that remain within the Earth-Moon sphere of influence (SOI) after 30 days, while gray markers indicate escaped fragments.}
  \label{fig:drotraj}
\end{figure*}

This section presents the dynamical evolution of debris clouds generated from spacecraft breakups across three distinct sizes of Distant Retrograde Orbits over a 30-day propagation period. Figure \ref{fig:drotraj} illustrates the spatial distribution of fragments. The purple markers denote the "trapped" fragments that remain within the Earth-Moon sphere of influence (SOI) at the end of the simulation, whereas the gray markers visualize the escape pathways of escaped debris, primarily through the vicinity of the $L_1$ and $L_2$ Lagrange points.

\label{lastpage}

\end{document}